\documentclass[12pt]{article}

\usepackage{latexsym}
\usepackage{graphicx}
\usepackage{rotating}
\usepackage[normalem]{ulem}
\usepackage{hyperref}
\usepackage{amsmath,amssymb,amsfonts}
\usepackage{bm}
\usepackage{color}

\newcommand{\dxy}{d$_{xy}$ }

\newcommand{\dxz}{d$_{xz}$ }

\newcommand{\vo}{VO$_2$}

\newcommand{\m}{M$_1$ }

\newcommand{\fref}[1]{Fig.~\ref{#1}}

\usepackage{color}

\newcommand{\lyxmathsym}[1]{\ifmmode\begingroup\def\b@ld{bold}
  \text{\ifx\math@version\b@ld\bfseries\fi#1}\endgroup\else#1\fi}

\usepackage{amsfonts}
\usepackage{mathrsfs}
\usepackage[latin9]{inputenc}
\usepackage{color}
\definecolor{note_fontcolor}{rgb}{0.800781, 0.800781, 0.800781}
\usepackage{mathrsfs}
\usepackage{amsmath}
\usepackage{amssymb}
\usepackage{graphicx}
\hypersetup{colorlinks=true,urlcolor=blue,linkcolor=blue}

\usepackage{scicite}

\usepackage{times}

\newenvironment{sciabstract}{%
\begin{quote} \bf}
{\end{quote}}

\title{Possible phonon-induced electronic bi-stability in VO$_2$ for ultrafast memory at room temperature}

\author
{C\'edric Weber,$^{1}$ Swagata Acharya,$^{1}$ \\ 
Brian Cunningham $^{2}$, Myrta Gr\"uning $^{2,3}$, Liangliang Zhang$^{4}$ \\
Hang Zhao$^{5}$, Yong Tan$^{5}$, Yan Zhang$^{4}$, Cunlin Zhang$^{4}$, Kai Liu$^{6}$,\\ 
Mark Van Schilfgaarde,$^{1}$ Mostafa Shalaby$^{4}$ \\
\\
\normalsize{$^{1}$King's College London, The Strand, WC2R 2LS London, UK,}\\
\normalsize{$^{2}$School of Mathematics and Physics, Queen's University Belfast, Belfast BT7 1NN, UK,}\\
\normalsize{$^{3}$European Theoretical Spectroscopy Facility (ETSF),}\\
\normalsize{$^{4}$Key Laboratory of Terahertz Optoelectronics,
Beijing Advanced}\\
\normalsize{Innovation Center for Imaging Technology CNU, Beijing 100048, China}\\
\normalsize{$^{5}$Beijing Key Laboratory for Precision Optoelectronic Measurement Instrument}\\
\normalsize{and Technology, School of Optics and Photonics, }\\
\normalsize{Beijing Institute of Technology, Beijing 100081, China}\\
\normalsize{$^{6}$School of Materials Science and Engineering,}\\
\normalsize{Tsinghua University, Beijing 100084, China,}\\
\\
}

\date{}

\begin{document}

\baselineskip24pt

\maketitle

\begin{sciabstract}
  VO$_{2}$ is a model material system which exhibits a metal to insulator transition at 67$^\circ$C. This holds potential for future ultrafast switching in memory devices, but typically requires a purely electronic process to avoid the slow lattice response. The role of lattice vibrations is thus important, but it is not well understood and it has been a long-standing source of controversy. We use a combination of ultrafast spectroscopy and ab initio quantum calculations to unveil the mechanism responsible for the transition. We identify an atypical Peierls vibrational mode which acts as a trigger for the transition. This rules out the long standing paradigm of a purely electronic Mott transition in \vo; however, we found a new electron-phonon pathway for a purely reversible electronic transition in a true bi-stable fashion under specific conditions. This 
transition is very atypical, as it involves purely charge-like excitations and requires only small nuclear displacement. Our findings will prompt the design of 
future ultrafast electro-resistive non-volatile memory devices.
\end{sciabstract}

\section*{Introduction}

Optical bi-stability is a key property to achieve ultrafast logic memory and switching devices \cite{liu_nat,li_mater_chem}.
 It is typically obtained when two distinct but nearly degenerate electronic states coexist,
and is controlled by  some combination of electronic charge and spin, and  their coupling to quantized lattice vibrations (phonons).
Its application to binary data storage is determined by the transition and relaxation times between these electronic states.
Bi-stability has been well studied in the context of magnetic systems, where the spin symmetry provides nearly degenerate ground states.
In these systems, the two stable states depend on the internal energy components of the material, such as the magnetic anisotropy, the exchange interactions, and applied external magnetic field.
However,  little progress has been made in finding bi-stable states involving charge degrees of freedom. 
Of the three degrees of freedom, charge, spin, and phonons, charge is the fastest and easiest to manipulate. At least, in principle,
it should offer a more versatile alternative to spin bi-stability in terms of speed and application.

However, ultrafast bi-stability in electronic systems is difficult to achieve in non-structured materials. Some materials show hysteretic response of the electron conductivity, for example under thermal excitations. However, only one state is strictly stable. The other  can be excited by an external stimulus and relaxes back after the stimulus is gone, and is thus not suitable for ultrafast memory applications. In terms of speed, while transient time to an excited state can occur on the sub-picosecond time scale, there is typically no way to trigger the reverse process  \cite{speed_limit_transition}.
 
Correlated electron systems offer various pathways of controlling material properties. For example, Vanadium dioxide (\vo) is an insulator with strong electron-phonon interactions and it undergoes a first order insulator-to-metal transition (IMT) at 340 K \cite{vo2_paper_ref_MIT}. At higher temperatures, \vo~ is metallic with the rutile structure (R), while it transforms to the monoclinic \m phase and becomes insulating below the transition temperature.

IMT can be achieved on the ultrafast time scale upon laser excitation \cite{dyn_cor_mater_1,dyn_cor_mater_2,dyn_cor_mater_3,dyn_cor_mater_4}. However, bi-stability is normally not reached as the
system undergoes an hysteresis. Whether the transition involves both electrons and phonons, or only electrons, remains a
topic of high controversy. The involvement of the lattice is detrimental for ultra-fast switching, as it is typically
associated with slow response and high power consumption. In this context, the lattice acts as a dissipation bath
(thermal reservoir) that prevents a reversible transition between the metallic and insulating phases \cite{Cavalleri_pump, kubler_pump,team_pump}. Moreover,
the nature of the MIT in \vo~ has long been debated, with particular emphasis placed on the role of electron
correlations in forming the charge gap. A key unanswered question is whether the IMT is driven by a pure electronic mechanism, as in the Mott transition, or if the Vanadium dimer pairing mechanism driven by Peierls distortions is responsible for the opening of the charge gap \cite{vo2_pump2,vo2_pump3,vo2_pump4,vo2_pump5}. This is paramount for future applications, such as low consumption non-volatile memory devices.


\section*{Discussion}

We  use a combination of ultra-fast terahertz (THz) spectroscopy techniques and quasi-particle self-consistent \emph{GW} theory (QS\emph{GW}).
In the first part, we address the controversy on whether the IMT in \vo~ is purely electronic, or is driven by lattice distortions. 

\begin{figure}
\begin{center}
\includegraphics[width=0.55\columnwidth]{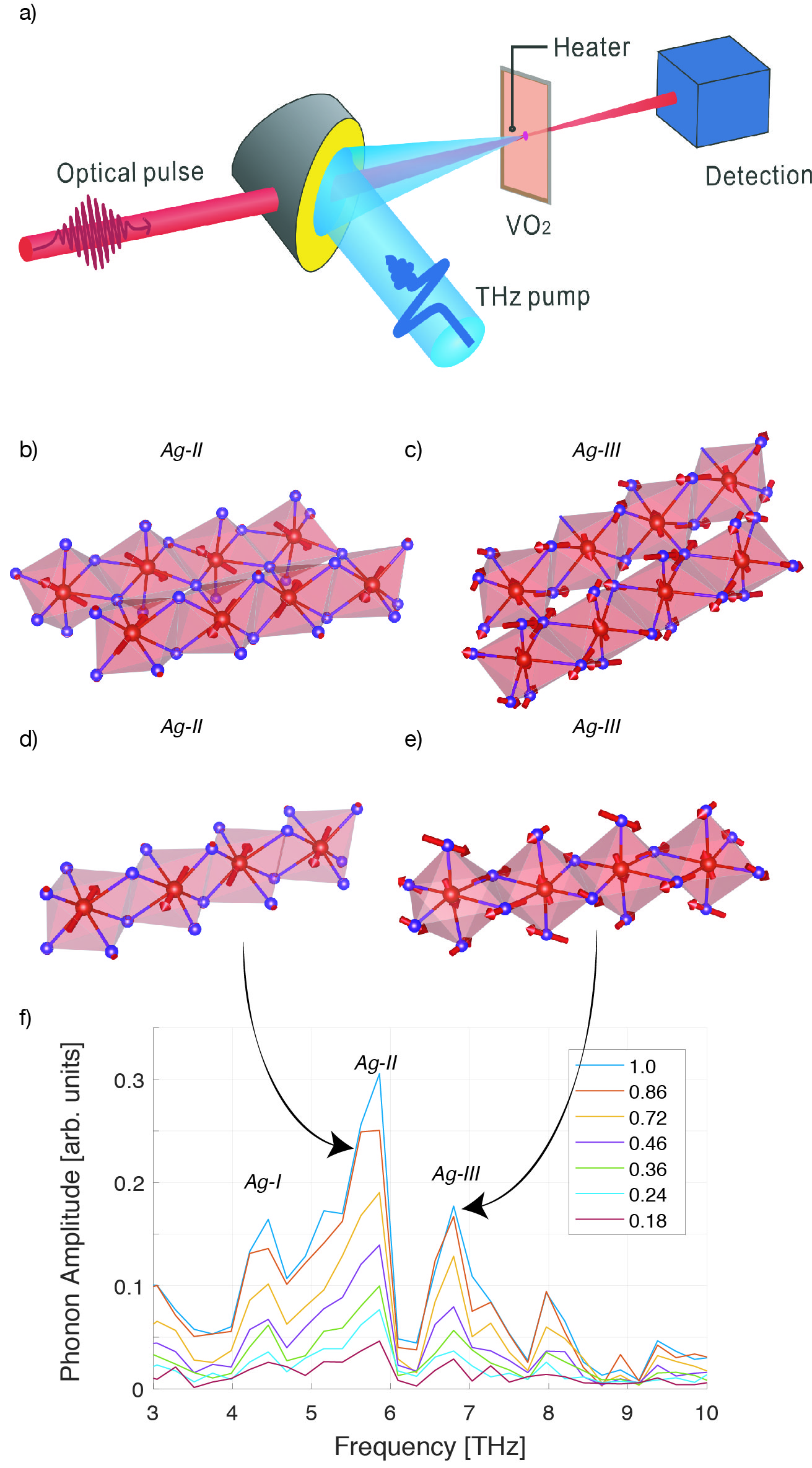}
\caption{
{\bf THz induced lattice distortions far from $T_c$ in \vo:}
a) The THz pump and optical probe setup.
Lattice distortions obtained in the calculated phonon eigenvector modes for b) the 5.7 THz A$_{g}$-II mode and c) the 6.8 THz A$_{g}$-III mode. Panel d) and e) show the enlargement of panel b) and c), respectively, of the V (red spheres) and O (blue spheres) edge-sharing octahedras along the rutile direction. The V-V dimers are located in the center of the rutile chain (second and third positions from the left).
f) The amplitude of phonon excitations for different terahertz power levels. The power is normalized by the largest intensity considered in the experiment.
}
\label{fig_summary}
\end{center}
\end{figure}

Structural properties are usually probed in equilibrium, and for \vo~ typically at temperatures near $T_{c}$. In our experiment, we use 800 nm probe to investigate the ultrafast dynamics immediately following generation of phonons by a THz pump. In this way the dominant excitations well
below $T_c$ can be observed.  A time resolution below 50\,fs is essential to reveal the timescales on which the excited electronic states redistribute their excess energy by coupling with phonons, and in turn can unveil the fingerprint (if any) of specific collective modes associated with the ultrafast rearrangement of the lattice. We performed two different time-resolved measurements: THz-pump, optical probe ellipsometry and THz-pump, optical probe transmission measurements (see \fref{fig_summary}). Both can reveal electron densities and phonon oscillations. However, ellipsometry is more sensitive to phonon oscillations (see \fref{fig_summary}) and transmission is very sensitive to the electron density. Our sample is a 70 nm-thick VO$_{2}$ film on a sapphire substrate.

\begin{figure}
\begin{center}
\hspace*{-0.4in}
\includegraphics[width=1.2\columnwidth]{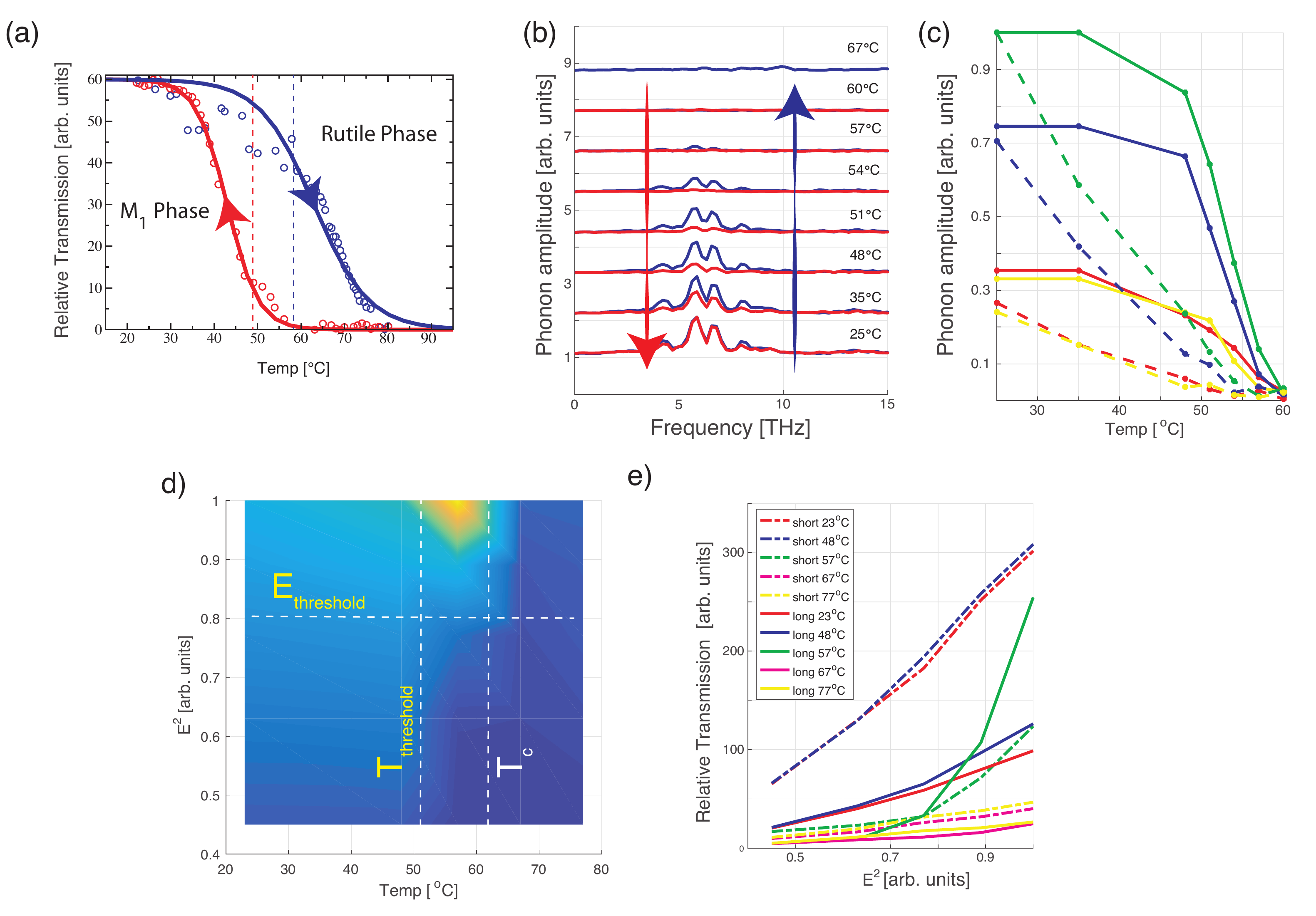}
\caption{ { \bf Ultrafast THz spectroscopy of electron and phonon excitations:}  
 a) Average transmission of the 800nm
probe obtained during the heating (blue) and cooling experiments (red).
b) Phonon amplitudes obtained by Fourier transform of the time-resolved THz pump optical ellipsometry measurements, at different temperatures between $20^\circ$C up to $67^\circ$C, normalised by the 5.7 THz mode obtained at $25^\circ$C. 
c) Hysteresis of the phonon amplitudes obtained in b).
(d) Phase diagram of the E-field dependent electronic response  obtained by time resolved THz pump and optical probe transmission at different temperatures (dark blue for low response, and yellow for high response). 
At temperature near Tc, $T > T_{threshold}$, and at high electric field $E > E_{threshold}$, the spectra is dominated 
by a sudden metallization of \vo\ via direct Zener-Keldysh-like interband tunneling (ZKIT). 
e) Power dependence of the 800 nm transmission at two delay points, respectively instantaneous (short) and at 20 ps (long).}
\label{exp3}
\end{center}
\end{figure}

We first show the spectroscopic ellipsometry measurements carried out along the rutile axis at 800\,nm (1.55\,eV),
 for temperatures ranging  between room temperature and the transition at 67$^\circ$C. In \fref{exp3}.a we 
 show the measured average temperature-transmission hysteresis. As expected, a hysteresis in the transmission is observed, which correlates with the first order transition of \vo~ at 67$^\circ$C.  However, the temporal response is highly non trivial.  In \fref{exp3}.b the birefringence on the optical probe is shown.  For all temperatures below 57$^\circ$C, a coherent temporal response is observed, with three dominant modes at  4.8 THz, 5.7 THz and
6.8 THz (see \fref{exp3}.b).  These frequencies closely match known phonon modes in the M$_{1}$ phase near 5 THz
(see Table~1 in SI).

Interestingly, the amplitude of these modes also exhibit hysteresis (see \fref{exp3}.b): the modes survive upon
heating to 57$^\circ$C, but only reappear upon cooling from high temperature at 50$^\circ$C. The hysteresis of the excitations, however, correlates with the hysteresis obtained in the averaged transmission, which shows that the excitations are an
inherent property of the \m phase of \vo. This also rules out heating effects induced by the THz pulse.  The fact that
the excitations are present all the way up to the phase transition establishes that the collective excitations are
connected to the first order transition of \vo~ and the collapse of the \vo~ gap at $T_c$.
These excitations clearly suggest that the lattice dynamics play a key role for the MI transition, and that the mechanism is Peierls-like, in
the sense that changes in electronic states are driven by small nuclear displacements.

\emph{Pump-probe Fluence.}
A fluence analysis (see Fig.~2 in SI) shows the modes' amplitudes vary
linearly with the square of the electric field.  This establishes that modes are not coherently excited by the THz pump,
but involve secondary relaxation mechanisms associated with the coupling of electrons to active phonon modes.  With THz
spectroscopy, electronic relaxation does not need to obey optical selection rules, but can occur through phonon coupling. We
note that the dominant mode is at 5.7 THz.

\emph{Optics.} To benchmark the QS\emph{GW} approximation with the experiments, we compute the optical conductivity including ladder diagrams via a Bethe-Salpeter formalism \cite{Cunningham18}. We report in the Fig.~1.b of SI the theoretical optical
conductivity along the rutile axis, and the comparison with ellipsometry measurement\cite{vo2_paper_ref3} of $\sigma_x$ on a single crystal
of \vo.  Agreement is excellent up to 4\,eV, especially when considering the complexity of this material, and also the
variability between different measurements of $\sigma$. The optical conductivity is a stringent test of the quality of QS\emph{GW}, as it is a true \emph{ab initio} theory, free of adjustable parameters.  

\emph{Phonons}. We computed the phonon dynamical matrix at the DFT level (see Table~1 in SI). The theoretical phonon
spectrum exhibits three modes in the range 4.8 THz to 7 THz, all of A$_{g}$ symmetry (denoted here as A$_{g}$-I,
A$_{g}$-II and A$_{g}$-III).  The phonon frequencies are in excellent agreement with the modes observed by THz
spectroscopy, and the phonon eigenvectors supply the nuclear displacements for each normal mode.  The
A$_{g}$-II (A$_{g}$-I) involves only displacements of the vanadium (oxygen) atoms, whereas the A$_{g}$-III mode 
involves displacement of both V and O ions. 

\emph{Phase diagram}. In \fref{exp3}.d, we report the average transmission at different temperatures. The colour map of the E-field dependent electronic response indicates that, at $T=57^\circ$C near $T_c=67^\circ$C,
a rapid metallization is suddenly obtained at large electric field $E>E_{threshold}$ (yellow region). This is also confirmed by
the rapid increase of the transmission at large electric fields 
$E^2>0.7$ for $T=57^\circ$C (\fref{exp3}.e). 
This has been associated \cite{team_pump} to a threshold field when. the absorbed energy density is sufficient to metallize the system via a structural change. In this regime, the monoclinic metallic phase is a dynamic precursor to the rutile metallic state.
 \fref{exp3}.d provides a phase boundary in terms of temperatures and electric fields, above which structural changes drive a metallic state. We focus in our work on lower temperatures and electric field, far from the Zener-Keldysh-like interband tunneling (ZKIT) regime. 

\begin{figure*}
\begin{center}
\hspace*{-0.85in}
\includegraphics[width=1.26\columnwidth]{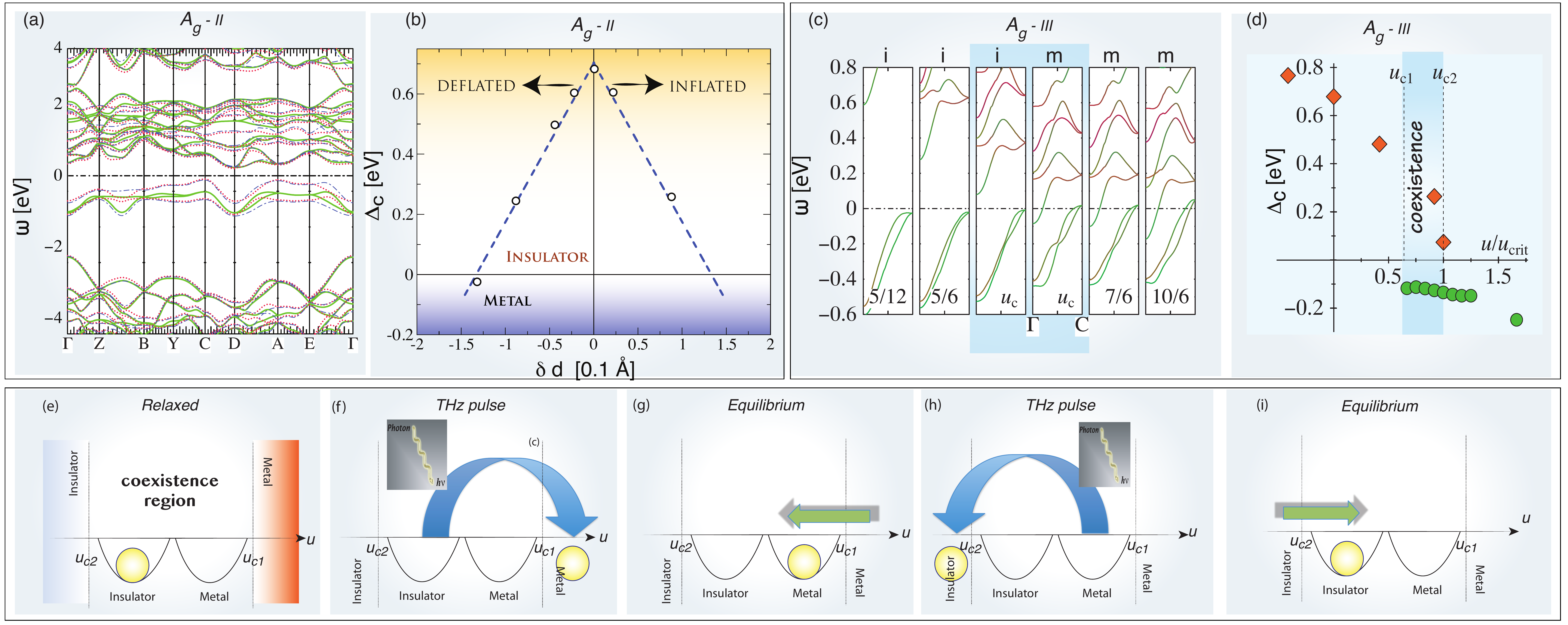}
\caption{
  {\bf Electronic bi-stability :} $a$) QS\emph{GW} energy band calculations for the displaced
  structure along the A$_{g}$-II (5.47 THz) phonon mode, with both expansions and contractions of the V-V bond
  ($\delta d \approx {\pm}0.022$\,\AA) (red and blue) around the M$_{1}$ phase (green). b) The charge gap $\Delta_c$ shrinks for either sign of $\delta d$ and eventually closes for changes in the V-V distance as small as $\delta d {\approx}
  0.15$\AA. 
  $c$) QS\emph{GW} energy band structure for the A$_{g}$-III (e.g., 6.29 THz) phonon mode for displacements above and below
  the critical displacement $u_c$, see text.  At $u_c$, QS\emph{GW} predicts the coexistence of two
  converged solutions: a metallic solution labelled \textbf{m} and an insulating solution labelled \textbf{i}.
  \emph{d}) Charge gap as a function of displacement $u$: we obtain an hysteresis and purely electronic IMT at $u_{c2}$ (diamonds) and respectively MIT at $u_{c1}$ (circles). 
  Bi-stability is obtained for $u_{c1}<u<u_{c2}$: if we surmise that the nuclear positions of the system
  can be pinned along the A$_{g}$-III mode in the coexistence region, so that the material does not relax to its pristine form, (e) bi-stability  would be achieved, and (f) an applied 6.29 THz pulse would drive the system to a metal.  After the system relaxed back to the
  middle region, (g) it would stay in the metallic solution, until (h) another pulse hits the system, after which (i) the
  system would relax to the insulating solution.}
\label{fig2}
\end{center}
\end{figure*}

\emph{Electron-phonon interaction}. We performed QS\emph{GW} calculations
in a frozen phonon approximation,
displacing atomic positions along eigenvectors of each of the three A$_{g}$ modes 4.81, 5.47, and 6.29 THz.  In two of these
modes (4.8 and 6.29 THz), a 2-fold rotation around the \emph{y} axis is preserved, but it is broken for the A$_{g}$-II mode (5.47 THz, see
Table~1 in SI).  \fref{fig2}.a shows how the quasiparticle band structure evolves for small displacements of
either sign ($\delta d \approx {\pm}0.022$\,\AA) in the A$_{g}$-II mode.  The nearly degenerate pair of $d$ bands just below $E_{F}$ split in a
symmetric manner, independent of the sign of displacement --- a characteristic signature of Peierls splitting --- while
the other bands are largely unperturbed.  As a result the gap decreases in proportion to $|\delta d|$ until it closes at
$\delta d\sim0.14$\,{\AA} (\fref{fig2}.b).  
Note that a negative $\Delta_c$ means that the solution is metallic: the conduction band at $\Gamma$ overlaps with the valence band, with emergence of electron and hole pockets.  
  
For comparison, a typical high power THz pump will displace ions by
roughly 0.1\,\AA, while nearest neighbor V-V bond lengths in the rutile and M$_{1}$ phase differ by $\sim$0.4\,\AA.
In the THz experiment, the ions are displaced in a complex manner, so a realistic simulation of the time-dependence
of $\sigma(\omega{=}1.5\mathrm{eV})$ is not feasible.  However, the conductivity $\sigma(\omega)$ is closely connected
to independent particle transitions (in particular the structure around 1.5\,eV is tied to transitions between the top two
valence band states and the unoccupied states), so the initial shape
of $\sigma(\omega)$ in the M$_{1}$ phase (see Fig.~1\emph{b} in SI) will evolve with excursions in the band structure
(Fig.~1\emph{a}S) from phonons generated by the pulse. This explains qualitatively the primary features of the THz experiment.

We next turn to the discussion of the A$_{g}$-III mode at 6.29 THz. This mode involves an approximately equal
displacements of the V and O ions, while preserving the two-fold rotational symmetry around \emph{y}.  The valence band
does not split and the charge gap evolves as a smooth function of phonon amplitude $u$, approaching a small positive
value at a critical value we denote as $u_{c}$ (\fref{fig2}.c). To provide some measure of the length scales involved,
at $u_{c}$ the nuclei are displaced by an average value of 0.06\,\AA, though the nearest neighbor V-V bond length increases by only
0.017\,\AA.  As $u$ increases slightly beyond $u_{c}$ the gap vanishes; the insulating solution becomes unstable
at $u_{c^\prime}{>}u_{c}$ and the
system makes a discontinuous transition to a metallic state, with a indirect negative gap of the order of $-0.15$\,eV, meaning
that the conduction band minimum (at $\Gamma$) falls slightly below the valence band maximum (which occurs at a
low-symmetry point roughly in the vicinity of C). \fref{fig2}.c displays the evolution of the bands along the
$\Gamma$-C line.

\emph{Bi-stability}. Remarkably, at $u_c$ two distinct self-consistent potentials can simultaneously be found \emph{for the same lattice
configuration}: one insulating and the other metallic. Both band structures are depicted in \fref{fig2}.c. This indicates
that a purely electronic transition is possible in this system.
The M-I transition here is very atypical, in that the associated fluctuations are purely charge-like, without involving the spin.
Note that the band structures of the coexisting solutions are similar; however there is a discontinuous change both in $\Delta_c$ and the valence band width. The narrowing of the bandwidth in the metallic solution indicates that the system is a correlated metal.
  
As a consequence of the coexistence of the two solutions, we obtain a hysteresis; with an excursion in the displacement
$u_{c}{<}u{<}u_{c^\prime}$ the system can remain on one branch or the other. The subtle balance between the localized
and itinerant character of the electrons in this regime is a realization of a \emph{spectral-weight scale}: if the
valence and conduction bands get slightly closer, the system gets metallic, if they move apart, the system becomes
insulating. Note that the M-I transition preserves the optical gap, as the charge gap is indirect.  This is distinct
from the insulator-metal transition in \vo~ at the critical temperature, where there is a collapse of the optical gap
concomitant with a collapse of the charge gap.
We note that this process occurs on the time scale of a phonon mode oscillation period, which is typically much faster than the
relaxation of resonant electronic excited modes. This opens a pathway for ultra-fast switching processes.

\emph{Doped compounds}. Finally, we extended the measurements to the case of doped $V_{1-x} W_x O_2$ with $x=0.01$. We observe that the Ag-III mode is still present upon chemically doping the material, with
a much reduced $T_c$, which opens promising avenues to further tuning the compounds towards the bi-stable phase (see text in SI).

\section{Conclusion}

We showed that a purely electronically driven transition does not occur for \vo~ at the critical temperature.
This concretely addresses a long-standing controversy on the role of phonons in the transition where the \m phase of \vo~ is a band insulator with a gap that is too large for pure many-body effects to stabilize a MIT without nuclear displacements.

Nevertheless, with suitable displacements relative to the M$_{1}$ phase, a purely electronic MI transition was found unveiling a new IMT mechanism: the Peierls instability which involves an orbital selection, and bonds the \dxy and \dxz orbitals along the rutile axis, filling each orbital with one electron. Nuclear displacements around the \m phase causes the gap to shrink.  When the gap becomes sufficiently small, a purely electronic MIT associated with the 6.8 THz phonon mode arises from the charge fluctuations.

This new mechanism enables bi-stability at room temperature in distorted structures which may be achieved by strain or external doping thus opening new possibilities for materials design under realistic experimental conditions.

\section*{Acknowledgments}
This work was supported by EPSRC (grants EP/R02992X/1, EP/N02396X/1, EP/M011631/1), and
the Simons Many-Electron Collaboration.
C.W. gratefully acknowledges the support of NVIDIA Corporation with the donation of the Tesla K40 GPUs used for this research. For computational resources, we were supported by the ARCHER UK National Supercomputing Service and the UK Materials and Molecular Modelling Hub for computational resources (EPSRC Grant No. EP/ P020194/1).


\clearpage

\clearpage

\clearpage

\clearpage
\clearpage

\end{document}